\documentclass[aps,pra,reprint,longbibliography, nofootinbib]{revtex4-2}
\usepackage[T1]{fontenc}
\usepackage[utf8]{inputenc}
\usepackage{lmodern}
\usepackage{microtype}
\usepackage{mathtools}
\usepackage{amsmath,amssymb,amsthm,amsfonts}
\usepackage{bm}
\usepackage{bbm}
\usepackage{xcolor}
\usepackage{overpic}
\usepackage[colorlinks = true, 
citecolor = blue,
linkcolor = blue,
urlcolor = blue]{hyperref}

\DeclareMathOperator{\tr}{Tr}
\DeclareMathOperator{\pr}{Pr}
\renewcommand{\Re}{\operatorname{Re}}
\renewcommand{\Im}{\operatorname{Im}}

\newcommand{\E}{\mathbb{E}}
\newcommand{\diff}{\,\mathrm{d}}

\newcommand{\coleq}{\mathrel{\mathop:}\nobreak\mkern-1.2mu=}

\usepackage{algorithm}  
\usepackage[noend]{algpseudocode}  

\DeclareMathOperator*{\argmax}{arg\,max}
\DeclareMathOperator*{\argmin}{arg\,min}

\begin{document}

\title{Quantum target ranging with Hetero-Homodyne detection}%

\author{Sangwoo Jeon}%
\altaffiliation[Current address: ]{Department of Combinatorics and Optimization and Institute for Quantum Computing, University of Waterloo, Waterloo, ON N2L 3G1, Canada}
\email{sangw077@gmail.com}

\author{Yonggi Jo}

\author{Jihwan Kim}

\author{Zaeill Kim}

\author{Duk Y. Kim}

\author{Yong Sup Ihn}

\author{Su-Yong Lee}
\email{suyong2@add.re.kr}

\affiliation{Agency for Defense Development, Daejeon 34186, Korea}

\begin{abstract}
Quantum target ranging, which estimates a target position using entangled photon pairs, is known to offer an error-probability advantage over classical ranging strategies.
Yet, realizing this advantage in practice remains challenging, as an existing receiver design relies on collective measurements and requires an impractically large number of quantum memories and linear passive components.
In this work, we propose a target ranging protocol with hetero-homodyne detection, providing a practically implementable architecture that achieves quantum advantage in target ranging using only local measurements.
Our protocol requires only one heterodyne setup, a single homodyne setup, a delay line, and an efficient classical postprocessing, making the implementation scalable and experimentally feasible.
Our results establish a realistic framework for demonstrating quantum advantage in target ranging and contribute toward practical quantum radar systems.

\end{abstract}
\maketitle
\section{Introduction}

Quantum illumination (QI)—a protocol that detects a low-reflectivity target at a specified range using entangled signal-idler states—has been shown to outperform classical illumination in regimes of weak returns and strong background noise, motivating quantum radar as a platform for practical quantum advantage~\cite{lloyd2008enhanced, tan2008quantum, shapiro2020quantum, sorelli2022detecting, torromé2024advances, karsa2024quantum}.
This has stimulated extensive efforts toward practical and implementable receiver designs for QI~\cite{guha2009gaussianstate, zhuang2017optimum, chang2019quantumenhanced, Jo2021, Yang21, lee2022observable, Lee23, reichert2023quantum, shi2024optimal, Jeon2025}, which ultimately enabled the experimental demonstration of its quantum advantage~\cite{Zhang2015, assouly2023quantum}.

Despite these advances, QI remains as a conceptual prototype of quantum radar, as it is limited to binary target detection, determining only whether a target is present or absent at a specified range.
By contrast, a fully developed quantum radar would require concrete information about the target, including its range, velocity, and structural features.
This limitation has motivated recent efforts to extend QI to target ranging—the task of estimating the position of a target~\cite{zhuang2021quantum, Karsa21, zhuang2022ultimate, liu2023entanglement, liao2024noisy, wang2025secure, khurana2025problem, Ortolano25}.
Notably, it has been established that quantum target ranging can, in principle, take an advantage over classical schemes~\cite{zhuang2021quantum}.

Naturally, proposing a practical target-ranging protocol stands as an important task. 
However, no feasible protocol achieving this advantage is currently known; the only existing protocol with quantum advantage relies on an impractically large number of optical components~\cite{liao2024noisy}. 
For instance, assuming a conservative input mode number of $10^5$, this would require on the order of $10^5$ quantum memories and $10^{10}$ programmable beam splitters, which is far beyond current technological capabilities. 
This impracticality arises from the requirement to simulate collective measurements over the received modes in order to attain the optimal performance. 
Therefore, it is natural to investigate suboptimal strategies that avoid collective measurements across copies and instead rely on local measurements performed on each returned mode. 
Such an approach has already proven effective in QI: while receivers based on collective measurements achieve the optimal quantum advantage at the cost of impractical resources~\cite{zhuang2017optimum, shi2024optimal}, simpler designs using local measurements can attain a suboptimal advantage~\cite{guha2009gaussianstate}, which has enabled a range of experimental demonstrations~\cite{Zhang2015, barzanjeh2020microwave, assouly2023quantum, Ward25}.

In this work, we propose a protocol for quantum target ranging that relies only on local measurements, a simple measurement architecture, and an efficient postprocessing while achieving quantum advantage.
This is the first quantum target ranging protocol that is both practically implementable and capable of demonstrating quantum advantage.

\section{Problem setup and backgrounds}

\begin{figure}[tb]
    \centering
    \includegraphics[width=0.86\linewidth]{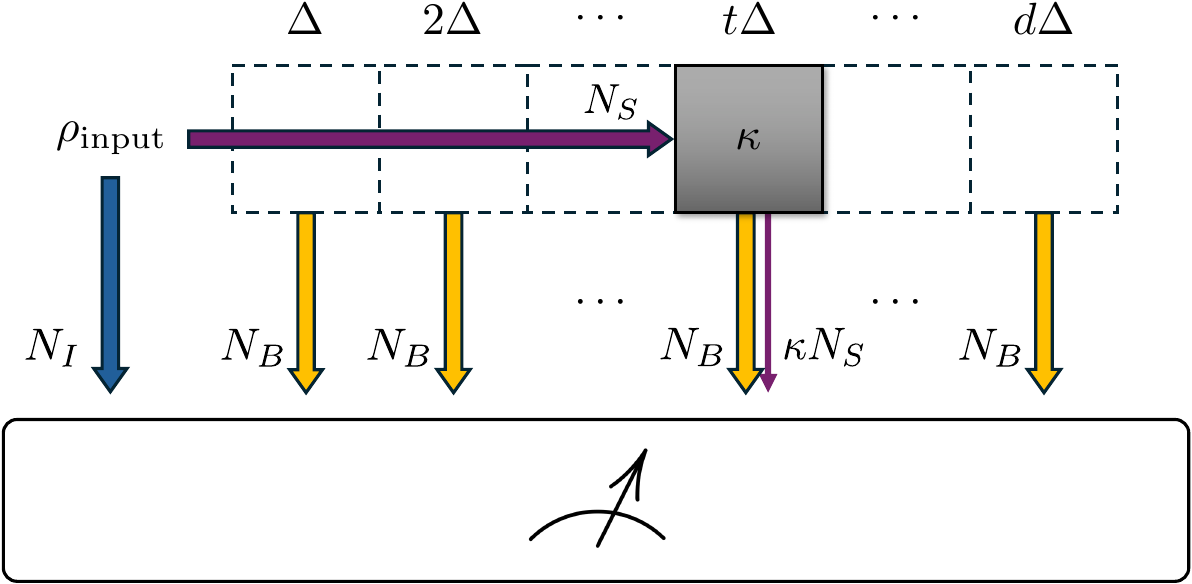}
    \caption{Schematic of the target ranging procedure. 
    A signal mode (purple) is transmitted to a one-dimensional target space while an entangled idler mode (blue) is retained. 
    The target space contains $d$ possible positions, and the receiver collects $d$ returned modes. 
    All returned modes are corrupted by thermal noise and become thermal states (yellow), whereas only the mode corresponding to the $t$-th position weakly retains information from the signal. 
    By measuring all returned modes together with the idler mode, the receiver estimates the target position.}
    \label{fig:schematic}
\end{figure}

We first describe the problem setup for target ranging and introduce the background. 
We consider target ranging as the task of estimating the position of a target using photon modes, which proceeds by transmitting a signal mode toward an estimated target position, measuring all returned modes, and estimating the target location through classical post-processing. 
An idler entangled with the signal is retained and measured jointly with the returned signal, as illustrated in Fig.~\ref{fig:schematic}. 
Throughout, we refer to quantum target ranging (QTR) as a protocol that employs an entangled idler and, by contrast, classical target ranging (CTR) as an idler-free protocol.\footnote{The idler-free model of CTR may not encompass all possible \textit{classical} protocols, as one may consider idler-assisted schemes with only classical correlations. For completeness, we show in Appendix~\ref{appsec:cct} that employing a classically correlated thermal state does not provide any advantage for target ranging.}

We formalize the target ranging task and introduce a suitable performance metric. 
Based on the settings in Refs.~\cite{zhuang2021quantum,liao2024noisy}, we adopt the following simplifications. 
In Fig.~\ref{fig:schematic}, first, we discretize the target space into a one-dimensional array, where possible target positions lie at distances $\Delta, 2\Delta, \dots, d\Delta$ from the source. 
Next, we discretize the photon modes by dividing the signal and idler into $M$ pulses.
Consequently, for each transmitted signal pulse, the receiver collects $d$ returned modes, with successive modes separated by a round-trip delay of $2\Delta/c$. 
As a result, for a target located at $t\Delta$ for some $t\in [d]$, all received modes are thermal states, and only the $t$-th mode weakly retains information from the returned signal. 
We model the target as a beam splitter with reflectivity $\kappa>0$, so that the returned modes are given by
\begin{align}
\hat{a}_{R,k}=
\begin{cases}
	\sqrt{\kappa}\hat{a}_{S}+\sqrt{1-\kappa}\hat{a}_{B} & k=t,\\
	\hat{a}_{B} & k\neq t,
\end{cases}
\end{align}
where $\hat{a}_{R,k}$ denotes the $k$-th returned mode, $\hat{a}_{S}$ the transmitted signal mode, and $\hat{a}_{B}$ a thermal noise mode. 
We set the mean photon number of the noise to $\langle\hat{a}_{B}^\dagger\hat{a}_{B}\rangle =N_B/(1-\kappa)$ for the reflected $t$-th mode and $\langle\hat{a}_{B}^\dagger\hat{a}_{B}\rangle =N_B$ for the other $k\neq t$, which allows us to ignore the shading effect of the target on the background noise, consistent with prior works on illumination and ranging~\cite{tan2008quantum, zhuang2021quantum}. 
Under these assumptions, the target ranging problem is converted to a multi-mode state discrimination task in which the receiver identifies which of the $d$ thermal modes retains information from the transmitted signal.

Further formalizing this task as a multiple-hypothesis testing problem allows us to consider an analytically tractable quantity—an error exponent—as a performance metric.
We restate the ranging task as a multiple-hypothesis testing problem
\begin{align}
H_1 \text{ vs. } H_2 \text{ vs. } \dots \text{ vs. } H_d,
\end{align}
where each hypothesis $H_k$ corresponds to the target being located at the $k$-th index. 
Under hypothesis $H_t$, the receiver obtains $M$ copies of the returned state $\rho_{\text{return},t}^{\otimes M}$ and applies a POVM $\{\hat{\Pi}_k\}_{k\in[d]}$ to infer the true target index $t$.
Assuming equal prior probabilities, the average error probability is then given by
\begin{align}
P_{\text{error}}=\frac{1}{d}\sum_{t=1}^d\text{Pr}(\text{Reject }H_{t}|H_t).
\end{align}
The quantum Chernoff bound (QCB) characterizes the asymptotic decay of the minimum achievable error probability in this setting.
Let $P_{\text{error, min}}$ denote the minimum $P_{\text{error}}$ over all possible POVMs.
Then, in the limit of large $M$, we have $P_{\text{error,min}}\sim \exp(-\xi M)$ with error exponent given by~\cite{li2016discriminating, nussbaum2011asymptotic, audenaert2007discriminating}
\begin{align}
\label{eq:exponent}
\xi = \min_{k\neq l}\max_{0\leq s\leq 1}\left[-\log\tr (\rho_{\text{return},k}^s\rho_{\text{return},l}^{1-s})\right].
\end{align}
Thus, comparing the error exponent in Eq.~\eqref{eq:exponent} allows us to assess the best achievable performance determined solely by the input states, independent of the specific measurement.

Under this setup, it is known that QTR can achieve a higher error exponent than any CTR scheme, thereby establishing a quantum advantage~\cite{zhuang2021quantum}. 
{To be specific, we consider QTR employing a two-mode squeezed vacuum (TMSV) state with signal mean photon number $\langle\hat{a}_S^\dagger\hat{a}_S\rangle=N_S$ and the same idler mean photon number $N_I=\langle\hat{a}_I^\dagger\hat{a}_I\rangle=N_S$, where $\hat{a}_I$ is the idler-mode annihilation operator, while CTR employs the best possible single-mode state with the same transmitted mean photon number $N_S$.}
In the parameter regime of $\kappa, N_S\ll 1 \ll N_B$, the corresponding error exponents are given by
\begin{align}
\xi_{\text{QTR}}&=\frac{2\kappa N_S}{N_B},\label{eq:qcb_qtr}\\
\xi_{\text{CTR}}&=\frac{\xi_{\text{QTR}}}{4},
\end{align}
which corresponds to a 6 dB quantum advantage of QTR over CTR.

It is important to note that this quantum advantage established in principle does not directly translate into a practical advantage, since the error exponent itself does not specify a ranging protocol that is physically implementable. 
Indeed, achieving quantum advantage in QTR remains challenging, as only impractical receiver designs have been proposed so far~\cite{shi2024optimal}. 
By contrast, for CTR, the QCB can be attained by transmitting coherent-state signals and performing homodyne measurements on the returned modes~\cite{zhuang2021quantum}.

\section{QTR protocol with Hetero-homodyne receiver}

\begin{figure}[tb]
    \centerline{
        \begin{overpic}[width=0.86\linewidth]{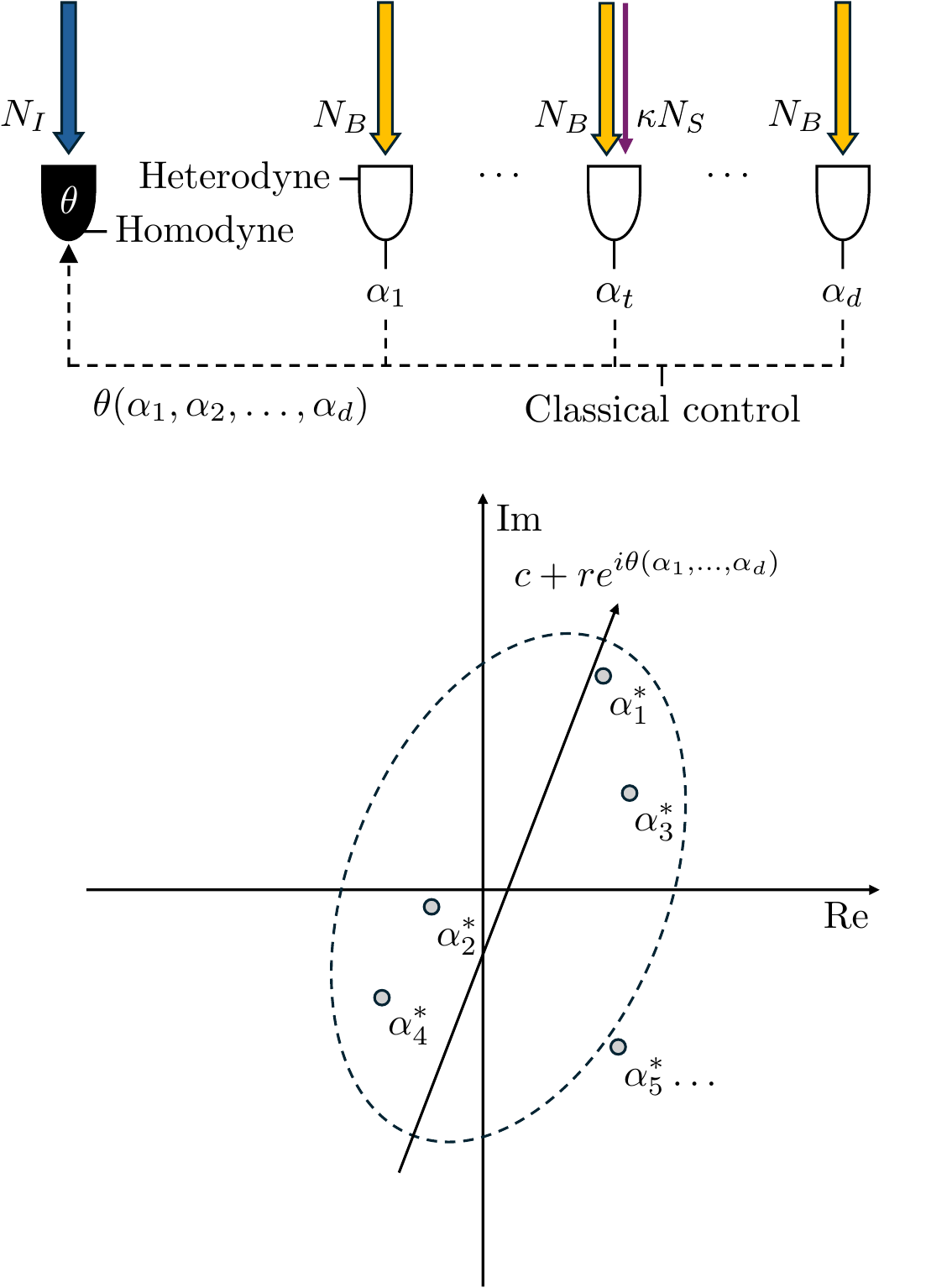}
        \put(-3,102){\text{(a)}} 
        \put(-3,63){\text{(b)}} 
    \end{overpic}
    }
    \caption{
    Schematic of the QTR protocol with HH receiver. 
    (a) Overall procedure. Heterodyne measurements are performed on the $d$ returned modes, and the outcomes are used to compute the measurement angle $\theta$, which sets the homodyne measurement basis. 
    The procedure is repeated over $M$ pulses, and the target index is estimated using an ML estimator. 
    (b) Selection of the measurement angle. 
    The angle $\theta$ is chosen as the argument of the first principal component of the complex conjugates of the heterodyne outcomes in the complex plane. 
    In the illustration, $c$ denotes a fixed complex number and $r$ a real parameter.
    }
    \label{fig:receiver}
\end{figure}

We present a QTR protocol utilizing the hetero-homodyne (HH) receiver, which extends the HH receiver ~\cite{reichert2023quantum} designed for QI.
The overall procedure is illustrated in Fig.~\ref{fig:receiver}(a). 
Initially, a TMSV state is prepared, with one mode transmitted to the target as the signal and the other retained on the transmitter side as the idler.
Heterodyne measurements are then performed on the $d$ returned modes, while the idler is stored in a delay line until all heterodyne measurements are completed.
Afterward, the idler is measured by homodyne detection with measurement angle $\theta$, which is determined from the $d$ heterodyne outcomes.
Precisely, denoting the outcomes by $\alpha_1,\dots,\alpha_d\in\mathbb{C}$, the angle $\theta$ is chosen as the argument of the first principal component of their complex conjugate $\alpha_1^*,\dots,\alpha_d^*$ in the complex plane, as illustrated in Fig.~\ref{fig:receiver}(b).
This procedure is repeated over $M$ pulses, and the collection of heterodyne and homodyne outcomes is processed by a maximum-likelihood (ML) estimator to infer the target index.
The detailed steps of the receiver are summarized in Algorithm \ref{alg:alg}, and closed-form expressions for $\theta_l$ in terms of $\{\alpha_{k,l}\}_{k\in[d]}$ together with the explicit form of the ML estimator are provided in Appendix~\ref{appsec:details}.

\begin{figure}[h]
\begin{algorithm}[H]
    \caption{HH receiver for QTR}
    \label{alg:alg}
    \begin{algorithmic}[1]
      \Require $M$ TMSV states.
      \Ensure Decide $H_t$.
        \For{$l=1~\textbf{to}~M$}
            \State Prepare a TMSV state.
                \State Send the signal mode to the target space and store the idler mode in the delay line.
            \For{$k=1$~\textbf{to}~$d$}
                \State Wait $2\Delta/c$.
                \State Heterodyne measurement $\alpha_{k,l}$ on the returned mode from the target space.
            \EndFor
        \State Compute $\theta_l$ with $\{\alpha_{k,l}\}_{k\in[d]}$.
        \State Homodyne measurement $X_l$ of the operator $\hat{x}\cos\theta_l+\hat{p}\sin\theta_l$ on the idler mode.
        \EndFor
        \State\Return ML estimate of $t$ using $\{\alpha_{k,l}\}_{(k,l)\in[d]\times [M]}$ and $\{X_l\}_{l\in[M]}$.
    \end{algorithmic}
\end{algorithm}
\end{figure}

Our protocol achieves the error exponent given by
\begin{align}
    \xi_{\text{HH}}=\left(1+\frac{\mathrm{B}(d/2,1/2)}{2}\right)\xi_{\text{CTR}},
\end{align}
where $\mathrm{B}(\cdot,\cdot)$ denotes the Beta function.
In particular, the exponent admits the following expressions in the small- and large-$d$ regimes:
\begin{align}
    \xi_{\text{HH}}=
    \begin{cases} 
      2\xi_{\text{CTR}} & d=2, \\
      \left(1+\sqrt{\pi/2d}\right)\xi_{\text{CTR}} & d\gg 1.
    \end{cases}.
    \label{eq:hh_exp}
\end{align}
It follows directly that $\xi_{\text{HH}}>\xi_{\text{CTR}}$, \textit{i.e.}, the protocol achieves quantum advantage in the asymptotic limit of large $M$.
{Note that the HH error-exponent gain depends on the number of target positions $d$, whereas the optimal QTR-over-CTR exponent gain is 6 dB and independent of $d$.} 
A detailed derivation of this result is provided in Appendix \ref{appsec:exp}.

\begin{figure*}[t!]
    \centering
    \begin{overpic}[width=0.31\textwidth]{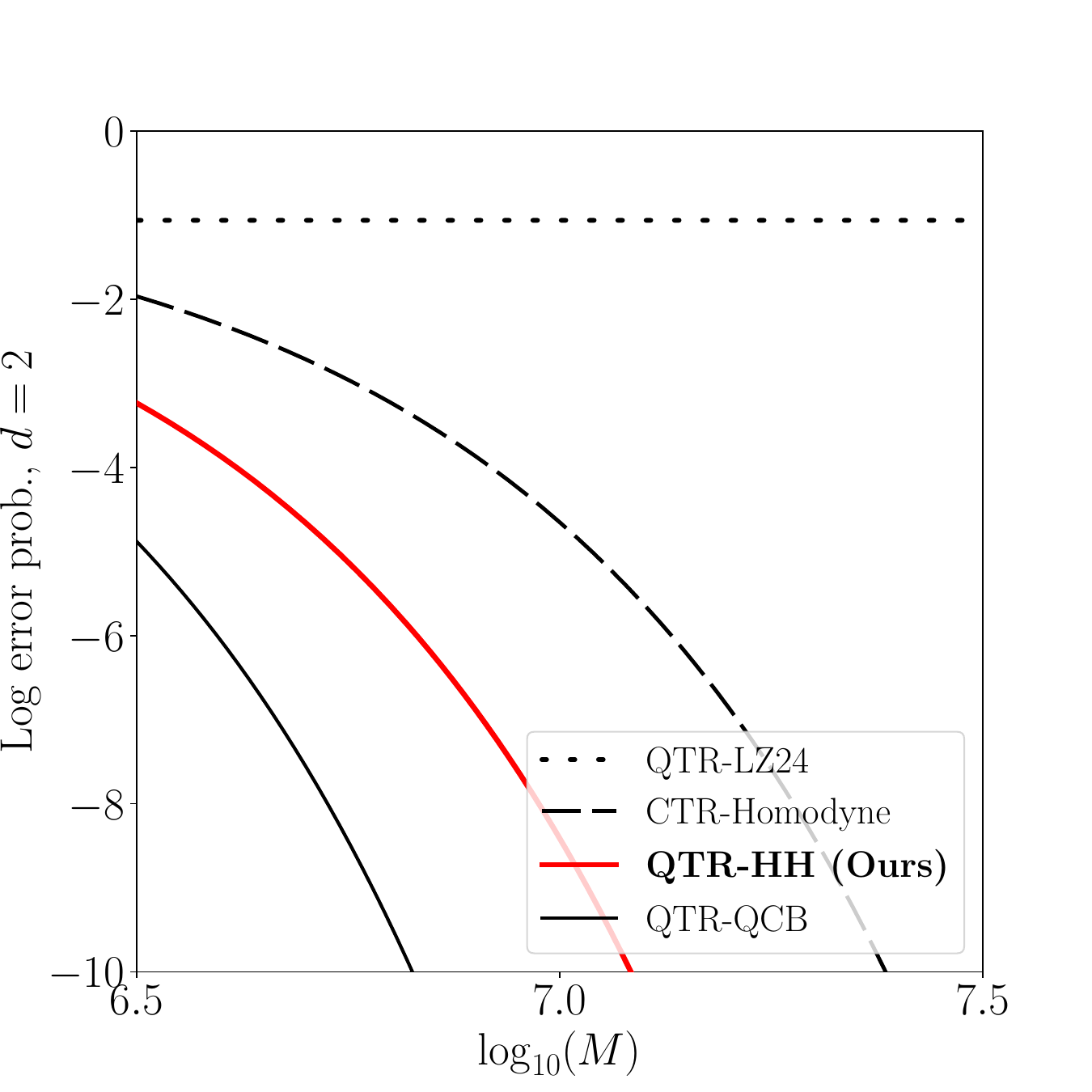}
        \put(0,89){(a)}
    \end{overpic}
    \hfill
    \begin{overpic}[width=0.31\textwidth]{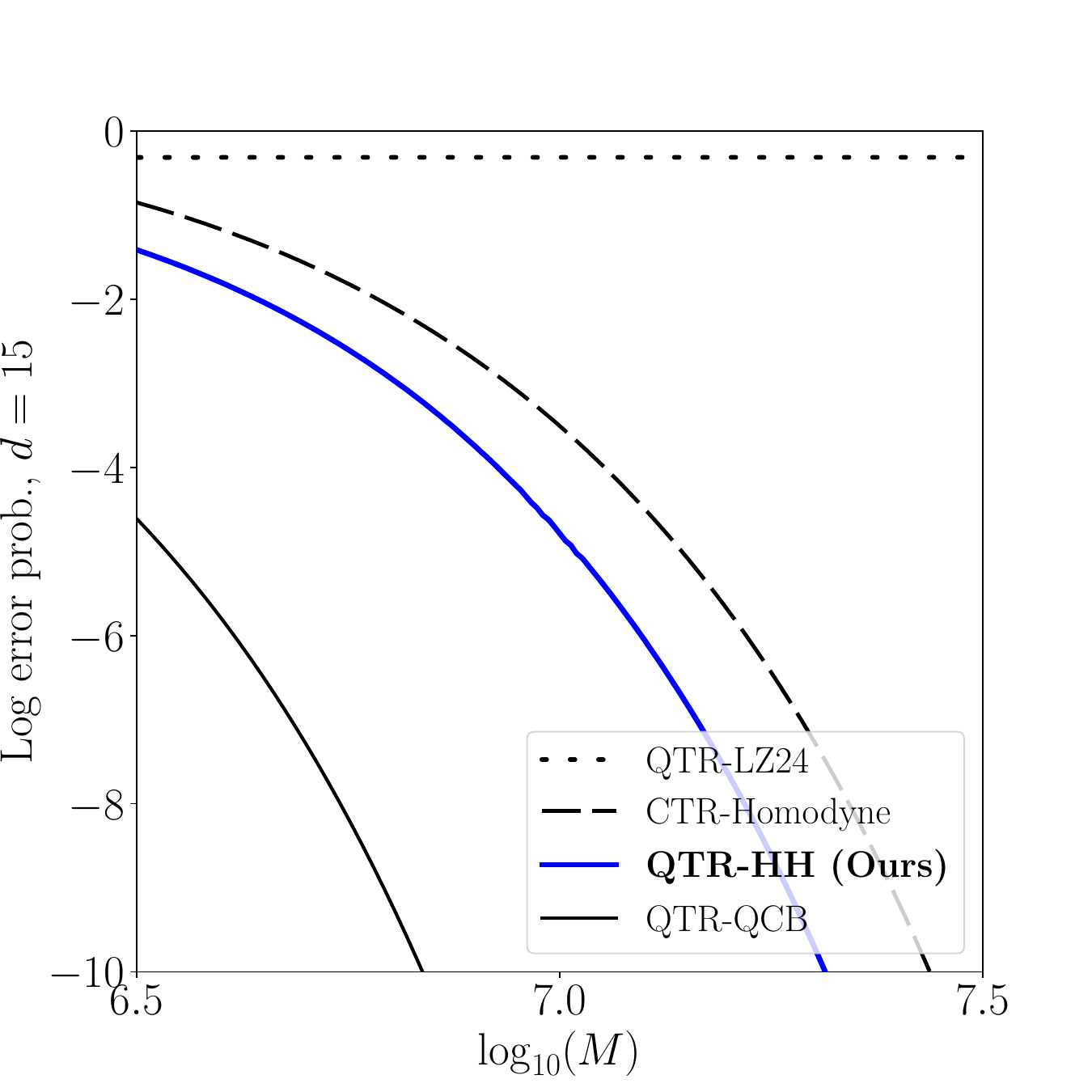}
        \put(0,89){(b)}
    \end{overpic}
    \hfill
    \begin{overpic}[width=0.31\textwidth]{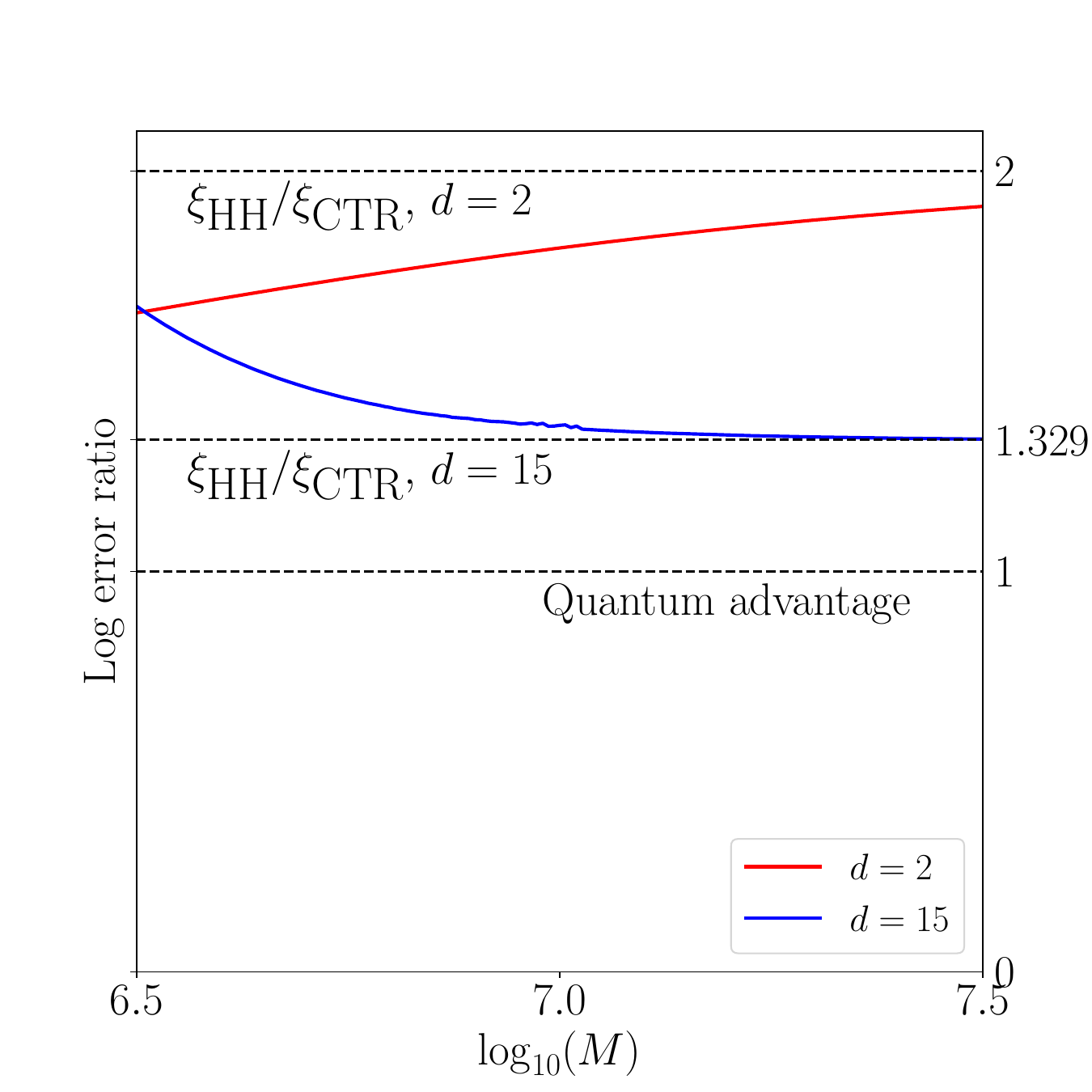}
        \put(0,89){(c)}
    \end{overpic}
    
    \caption{
        {
        Numerical simulations for $d=2$ and $d=15$ with parameters $N_B=600$, $\kappa=0.01$, and $N_S=0.1$.
        (a) Logarithms of the error probabilities for QTR and CTR as functions of $M$ for $d=2$, and (b) the corresponding results for $d=15$.
        QTR-LZ24 denotes the QTR receiver proposed in Ref.~\cite{liao2024noisy}.
        CTR-Homodyne denotes coherent-state transmission followed by homodyne detection, which attains the optimal error exponent of CTR in the asymptotic regime.
        QTR-HH denotes the proposed HH receiver.
        QTR-QCB denotes the QCB-based optimal-QTR benchmark obtained from Eq.~\eqref{eq:qcb_qtr}.
        (c) Ratio between the logarithms of the HH and CTR error probabilities.
        The dashed lines indicate, from top to bottom, the asymptotic error-exponent ratio $\xi_{\mathrm{HH}}/\xi_{\mathrm{CTR}}$ for $d=2$, the corresponding ratio for $d=15$, and the quantum-advantage threshold at ratio 1.
        }
    }

    \label{fig:numerical}
\end{figure*}

We emphasize that the HH receiver is experimentally feasible.
It requires only a single heterodyne and a single homodyne measurement setup, which makes the implementation straightforward and scalable in the number of target positions $d$ and the number of pulses $M$. 
Importantly, no active quantum memory is required, since the idler can be stored in a delay line for a fixed time until one round of heterodyne measurements is completed.
This is a clear improvement over the receiver of Ref.~\cite{liao2024noisy}, which requires storing $M$ idler modes in $M$ quantum memories and retrieving them in precise synchrony. 
We also note that the classical postprocessing is efficient as well.
For each pulse, computing $\theta$ requires forming a $2\times 2$ covariance matrix from $d$ heterodyne outcomes, which takes $\mathcal{O}(d)$ time. 
The ML estimator then evaluates $d$ squared norms of $M$-dimensional vectors, leading to an overall runtime of $\mathcal{O}(dM)$, which remains well within experimentally practical limits.

The key intuition behind the quantum advantage of our protocol lies in the choice of the homodyne measurement angle. 
After the heterodyne measurements are completed for a given TMSV pulse, the idler mode becomes a displaced thermal state whose displacement depends on the heterodyne outcome at the true target index, $\alpha_t$~\cite{shi2024optimal, reichert2023quantum, liao2024noisy}. 
More precisely, the displacement is proportional to the complex conjugate of the outcome and can be written as $c\alpha_t^*$ for a fixed constant $c$. 
Consequently, the homodyne measurement on the idler must discriminate among the candidate displacements $c\alpha_1^*,\dots,c\alpha_d^*$. 
Since homodyne measurement accesses only the projection of the displacement along a single quadrature, the choice of measurement basis becomes crucial. 
By aligning the measurement basis with the first principal component of the heterodyne outcomes in the complex plane, the receiver enhances the separation between the candidate displacements, driving the error probability into a quantum-advantageous regime.
{This intuition explains the $d$-dependence in Eq.~\eqref{eq:hh_exp}: as $d$ increases, the scatter ellipse in Fig.~\ref{fig:receiver}(b) becomes less elongated, making the alignment to the principal component less beneficial and thereby reducing the quantum advantage.}

We numerically validate our theoretical analysis in Fig.~\ref{fig:numerical}.
{
Figs.~\ref{fig:numerical}(a) and \ref{fig:numerical}(b) compare the log error probabilities of the HH receiver for $d=2$ and $d=15$, respectively, with several benchmarks: the QTR receiver proposed in Ref.~\cite{liao2024noisy}, coherent-state CTR with homodyne detection, and the optimal-QTR benchmark obtained from QCB.
The results show that the HH receiver outperforms coherent-state CTR with homodyne detection throughout the plotted range, demonstrating quantum advantage in a realistic parameter regime, including the case of large $d$.
Notably, for the considered parameters, the HH receiver also outperforms the receiver of Ref.~\cite{liao2024noisy}, whose error probability saturates to a constant.
This saturation originates from the conditional-nulling step of the receiver, which yields a nonzero false-alarm probability that scales linearly with the signal mean photon number $N_S$ to leading order~\cite{liao2024noisy}.
As a result, the receiver of Ref.~\cite{liao2024noisy} attains the optimal error exponent only in the low-brightness limit $N_S\to 0$.
Thus, the HH receiver can offer an additional practical advantage over the previously known receiver by achieving quantum advantage at finite $N_S$.
Fig.~\ref{fig:numerical}(c) presents the ratio between the logarithms of the HH and CTR error probabilities, providing a direct visualization of the quantum advantage achieved by the HH receiver.
Note that the apparent overlap between the $d=2$ and $d=15$ curves in the smaller-$M$ regime arises from $d$-dependent nonasymptotic prefactors in the HH and CTR error probabilities.
}

\section{Discussions}

In this work, we introduced the QTR protocol utilizing the HH receiver and demonstrated that it achieves quantum advantage over classical schemes while remaining experimentally feasible. 
Our protocol requires only a single heterodyne setup, a single homodyne setup, a delay line, and simple classical information processing, thereby avoiding the demanding resources required by the previously proposed receiver.
This establishes our protocol as a practical platform for experimentally demonstrating quantum advantage in ranging tasks.
Beyond its practical simplicity, our results show that collective entangled measurements across copies are not necessary to realize a quantum advantage in target ranging.

We note that the HH receiver may offer an additional practical advantage in continuous-wave implementations. 
Although our analysis is based on a discretized pulse model, realistic target ranging systems typically operate in a continuous-wave regime. 
In such settings, coherent-state schemes generally require external modulation or phase coding to encode timing information, since an unmodulated coherent beam alone does not provide ranging capability.
By contrast, the HH receiver does not require external modulation. 
The heterodyne outcomes, which are random for each shot, naturally provide displacement references for the homodyne measurement of the idler, enabling range information to be extracted directly. 
This suggests that the HH receiver could achieve additional resource efficiency in continuous-wave operation. 
We emphasize, however, that realistic implementations may introduce correlations between heterodyne outcomes depending on the signal and measurement bandwidth, and analyzing these effects remains an important direction for future work.

We conclude by outlining several directions for future work. 
Since the TMSV state is known to achieve optimal performance in covert sensing, extending the HH receiver to covert ranging constitutes a natural next step~\cite{tham2024quantum}. 
Another promising direction is to further optimize the choice of the measurement angle $\theta$. 
For example, adaptively selecting $\theta$ based on heterodyne outcomes accumulated from previous pulses may further improve the receiver performance. 
{Furthermore, we may think about how to characterize the optimal quantum advantage achievable with local measurements and whether this advantage exhibits a $d$-dependence as observed by the HH receiver.}

\begin{acknowledgments}
This work was supported by Defense Acquisition Program Administration and Agency for Defense Development.
\end{acknowledgments}

\bibliography{references.bib}

\begingroup
\onecolumngrid

\appendix
\section{Conventions for continuous-variable systems}

We summarize the conventions utilized throughout the paper.
The operator conventions $\hat{x}\coleq\hat{a}+\hat{a}^\dagger$ and $\hat{p}\coleq(\hat{a}-\hat{a}^\dagger)/i$ are employed, and the covariance matrix of a Gaussian state is defined as $V\coleq[\langle\{{r}_i,{r}_j\}/2\rangle]_{i,j}$ for a vector $\boldsymbol{r}=(r_1,r_2,\dots)^T\coleq(\hat{x}_1,\hat{p}_1,\dots)^T$, where $\{\cdot,\cdot\}$ denotes the anti-commutator. 

Under this convention, the covariance matrix of the TMSV state with mean photon number $N_S$ is given by
\begin{align}
    V=\begin{pmatrix}
        (2N_S+1){I}_2 & 2\sqrt{N_S(N_S+1)}{Z}_2\\
        2\sqrt{N_S(N_S+1)}{Z}_2 & (2N_S+1){I}_2
    \end{pmatrix},
\end{align}
where $I_2$ is the $2\times 2$ identity matrix and $Z_2$ is the $2\times 2$ Pauli-$Z$ matrix. 
Note that all vectors are denoted in boldface.

\section{QCB of classically correlated thermal state input}
\label{appsec:cct}

We show that the error exponent of target ranging using a classically correlated thermal state cannot exceed $\xi_{\mathrm{CTR}}$. 
Consider a classically correlated thermal state produced by impinging a thermal state on a beam splitter, which has the covariance matrix~\cite{lee2022observable}
\begin{align}
    V_\mathrm{CCT}=\begin{pmatrix}
        (2N_S+1){I}_2 & 2\sqrt{N_S N_I}{I}_2\\
        2\sqrt{N_S N_I}{I}_2 & (2N_I+1){I}_2
    \end{pmatrix}.
\end{align}
The covariance matrix of the joint state of the $t$-th returned mode and the idler mode under $H_t$ is then given by
\begin{align}
    V_{t,I}=\begin{pmatrix}
        (2N_B+2\kappa N_S+1){I}_2 & 2\sqrt{\kappa N_S N_I}{I}_2\\
        2\sqrt{\kappa N_S N_I}{I}_2 & (2N_I+1){I}_2
    \end{pmatrix}.
\end{align}

The error exponent of the multiple-hypothesis task can be reduced to that of a binary hypothesis test~\cite{li2016discriminating, nussbaum2011asymptotic}. 
More precisely, the error exponent of the target ranging task reduces to that of discriminating between two states with the following covariance matrices:
\begin{align}
    V_{t,I}^{(0)}&=\begin{pmatrix}
        (2N_B+2\kappa N_S+1){I}_2 & 0 & 2\sqrt{\kappa N_S N_I}{I}_2\\
        0 & (2N_B+1){I}_2 & 0\\
        2\sqrt{\kappa N_S N_I}{I}_2 & 0 & (2N_I+1){I}_2
    \end{pmatrix},\\
    V_{t,I}^{(1)}&=\begin{pmatrix}
        (2N_B+1){I}_2 & 0 & 0\\
        0 & (2N_B+2\kappa N_S+1){I}_2 & 2\sqrt{\kappa N_S N_I}{I}_2\\
        0 & 2\sqrt{\kappa N_S N_I}{I}_2 & (2N_I+1){I}_2
    \end{pmatrix}.
\end{align}

This can be evaluated analytically using the methods of Ref.~\cite{pirandola2008computable}. 
Although the full derivation involves lengthy calculations, an approximation under $N_S,\kappa\ll 1\ll N_B, N_I$ yields the leading-order term of the error exponent,
\begin{align}
    \xi_\mathrm{CCT}=\xi_{\mathrm{CTR }}=\frac{\kappa N_S}{2N_B},
\end{align}
which only achieves the classical bound.
Under the regime of $N_S=N_I\ll 1$, the error exponent reduces to
\begin{align}
    \xi_\mathrm{CCT}=\frac{2\kappa N_S^2}{N_B},
\end{align}
which is strictly smaller than the classical limit.

\section{Homodyne angle and ML estimator}
\label{appsec:details}

We provide closed-form expressions for the homodyne measurement angle and the ML estimator.

\subsection{Homodyne angle}

Let $\alpha_{1}^*,\dots,\alpha_d^*$ denote the heterodyne measurement outcomes and $\theta$ the resulting measurement angle. 
Define $\mu=\frac{1}{d}\sum_{k=1}^d\alpha_k^*$ and $z_k=\alpha_k^*-\mu$ for all $k\in[d]$. 
Writing $S_{xx}=\sum_{k=1}^d(\Re(z_k))^2$, $S_{yy}=\sum_{k=1}^d(\Im(z_k))^2$, and $S_{xy}=\sum_{k=1}^d\Re(z_k)\Im(z_k)$, the vector $\boldsymbol{v}=(\cos\theta,\sin\theta)^T$ is the first principal component of the points $\{(\Re(z_k), \Im(z_k))\}_{k\in[d]}$, \textit{i.e.}, the eigenvector with the largest eigenvalue of the covariance matrix $\Sigma$, defined as
\begin{align}
    \Sigma=
    \begin{bmatrix}
        S_{xx} & S_{xy}\\
        S_{xy} & S_{yy}
    \end{bmatrix}.
\end{align}
Thus, finding $\theta$ that maximizes $\boldsymbol{v}^T\Sigma\boldsymbol{v}$ yields the measurement angle. 
We have
\begin{align}
    \boldsymbol{v}^T\Sigma\boldsymbol{v}
    &= S_{xx}\cos^2\theta+2S_{xy}\cos\theta\sin\theta+S_{yy}\sin^2\theta\\
    &=\frac{S_{xx}+S_{yy}}{2}+\frac{S_{xx}-S_{yy}}{2}\cos2\theta+S_{xy}\sin2\theta,
\end{align}
which is maximized at
\begin{align}
    \theta=\frac{1}{2}\tan^{-1}\left(\frac{2S_{xy}}{S_{xx}-S_{yy}}\right).
\end{align}
Equivalently, the measurement angle can be written as the argument of a complex number,
\begin{align}
    \theta
    &=\frac{1}{2}\arg\left(\sum_{k=1}^d(\alpha_k^*-\mu)^2\right).
\end{align}

\subsection{ML estimator}

Let the heterodyne measurement outcomes be $\boldsymbol{\alpha}_k\coleq(\alpha_{k,1},\dots,\alpha_{k,M})^T$ for $k\in[d]$, the homodyne measurement outcomes be $\boldsymbol{X}\coleq(X_1,\dots,X_M)^T$, and the homodyne measurement angles be $\boldsymbol{\theta}\coleq(\theta_1,\dots,\theta_M)^T$. 
Under $H_k$, the homodyne outcomes $\boldsymbol{X}$ follow an $M$-dimensional standard normal distribution with mean $\boldsymbol{\mu}_k$ given by
\begin{align}
    \boldsymbol{\mu}_k=\frac{2\sqrt{\kappa N_S}}{N_B}\Re(\boldsymbol{\alpha}^*_k\circ e^{-i\boldsymbol{\theta}}),
\end{align}
where $\circ$ denotes the component-wise product (see Appendix~\ref{appsec:exp} for a proof). 
Thus, the ML estimator returns the index $\tilde{t}$ for which $\boldsymbol{\mu}_{\tilde{t}}$ is closest to $\boldsymbol{X}$ in Euclidean norm, \textit{i.e.},
\begin{align}
    \tilde{t}
    &=\argmin_k \|\boldsymbol{X}-\boldsymbol{\mu}_k\|\\
    &=\argmin_k \left\|\boldsymbol{X}-\frac{2\sqrt{\kappa N_S}}{N_B}\Re(\boldsymbol{\alpha}^*_k\circ e^{-i\boldsymbol{\theta}})\right\|.
\end{align}

\section{Error exponent of QTR}
\label{appsec:exp}

We derive the error exponent of our target-ranging protocol. 
Recall that the protocol consists of two steps: performing consecutive heterodyne and homodyne measurements, and then estimating the target range using an ML estimator. 
In the following, we first derive the outcome distributions of the heterodyne and homodyne measurements explicitly, and then obtain the error exponent of the maximum-likelihood estimator in closed form.

We derive the distributions of the observables obtained from the heterodyne and homodyne measurements. 
In the heterodyne stage, the receiver obtains the $d$ modes from the target space, where only the $t$-th returned mode weakly attains information from the signal. 
The covariance matrix of the joint state of the $t$-th returned mode and the idler is given by
\begin{align}
    V_{t,I}=\begin{pmatrix}
        (2N_B+2\kappa N_S+1){I}_2 & 2\sqrt{\kappa N_S(N_S+1)}{Z}_2\\
        2\sqrt{\kappa N_S(N_S+1)}{Z}_2 & (2N_S+1){I}_2
    \end{pmatrix},
\end{align}
whereas for indices $k\neq t$,
\begin{align}
    V_{k\neq t,I}=\begin{pmatrix}
        (2N_B+1){I}_2 & 0\\
        0 & (2N_S+1){I}_2
    \end{pmatrix}.
\end{align}
Thus, each heterodyne outcome from the received modes $\alpha_{1},\dots, \alpha_{d}\in\mathbb{C}$ follows a complex normal distribution
\begin{align}
    \alpha_{t}&\sim\mathcal{CN}\left(0,N_B+1\right),
\end{align}
where $\alpha\sim\mathcal{CN}(0,\sigma^2)$ means that the real and imaginary parts of $\alpha$ are independent and distributed as $\Re(\alpha),\Im(\alpha)\sim\mathcal{N}(0,\sigma^2/2)$.

Turning to the homodyne measurement, the idler mode conditioned on a heterodyne outcome becomes a displaced thermal state, with the displacement depending on the outcome $\alpha_t$ from the target. 
Explicitly, by taking the partial trace over the joint state of the reflected mode and the idler, the conditional idler state $\rho_{I|\alpha_t}$ is given by~\cite{shi2024optimal}
\begin{align}
    \rho_{I|\alpha_t}&\coleq\hat{D}({\mu}_t)\rho_{\text{th}}(N_{\text{th}})\hat{D}^\dagger({\mu}_t),\\
    {\mu}_t&\coleq\frac{\sqrt{\kappa N_S(N_S+1)}}{N_B+\kappa N_S+1}\alpha_t^*,\\
    N_{\text{th}}&\coleq\frac{N_S(N_B+1-\kappa)}{N_B+\kappa N_S+1}.
\end{align}
Here, $\hat{D}(\mu_t)$ is the displacement operator with displacement $\mu_t$, and $\rho_{\text{th}}(N_{\text{th}})$ is a thermal state with mean photon number $N_{\text{th}}$. 
The homodyne outcome $X$ for the operator $\hat{x}\cos\theta+\hat{p}\sin\theta$ at measurement angle $\theta$ is then distributed as
\begin{align}
    X\sim\mathcal{N}\left(2\Re(\mu_t e^{-i\theta}),2N_{\text{th}}+1\right).
\end{align}
Using the approximation
\begin{align}
    \mu_t&\approx\frac{\sqrt{\kappa N_S}}{N_B}\alpha_t^*,\\
    N_{\text{th}}&\approx N_S,
\end{align}
under $\kappa, N_S\ll1\ll N_B$, the measurement outcome distributions can be approximately rewritten as
\begin{align}
    \alpha_k&\sim\mathcal{CN}(0,N_B),\\
    X&\sim\mathcal{N}\left(\frac{2\sqrt{\kappa N_S}}{N_B}\Re(\alpha_t^* e^{-i\theta}),1\right).
\end{align}

Collecting these results, we can express the full measurement outcomes in a unified form. 
For the $l$-th signal–idler pulse, let $\alpha_{k,l}$ denote the heterodyne measurement outcome from the $k$-th returned mode, $X_l$ the homodyne measurement outcome, and $\theta_l$ the corresponding homodyne measurement angle. 
Define the vectors $\boldsymbol{\alpha}_k\coleq(\alpha_{k,1}, \dots, \alpha_{k,M})^T$ for $k\in[d]$, $\boldsymbol{X}\coleq(X_1,\dots,X_M)^T$, and $\boldsymbol{\theta}\coleq(\theta_1,\dots,\theta_M)^T$.
Then the measurement outcomes satisfy
\begin{align}
    \boldsymbol{\alpha}_k&\sim\mathcal{CN}\left(\boldsymbol{0}_M,N_B{I}_M\right),\\
    \boldsymbol{X}&\sim\mathcal{N}\left(\frac{2\sqrt{\kappa N_S}}{N_B}\Re(\boldsymbol{\alpha}_t^*\circ e^{-i\boldsymbol{\theta}}),I_M\right), \label{eq:Gaussian}
\end{align}
where $\circ$ denotes the component-wise product of two vectors and $I_M$ is the $M$-dimensional identity matrix.

Given that the obtained measurement outcomes $\boldsymbol{\alpha}_1,\dots,\boldsymbol{\alpha}_d,\boldsymbol{X}$ are all Gaussian random variables, we explicitly derive the error exponent from the ML estimator in our algorithm. 
The likelihood $\mathcal{L}(\boldsymbol{X};{\boldsymbol{\alpha}_k,\boldsymbol{\theta}})$ of hypothesis $H_k$ is directly obtained from Eq.~\eqref{eq:Gaussian} and takes the form of a Gaussian function
\begin{align}
    \mathcal{L}(\boldsymbol{X};{\boldsymbol{\alpha}_k,\boldsymbol{\theta}})\coleq\exp\left(-\frac{1}{2}\left\|\boldsymbol{X}-\frac{2\sqrt{\kappa N_S}}{N_B}\Re(\boldsymbol{\alpha}_k^*\circ e^{-i\boldsymbol{\theta}})\right\|^2\right),
\end{align}
where the constant prefactor is omitted.
Then, the error probability of the ML estimator is bounded as follows:
\begin{align}
    P_{\text{error}}
    &=\text{Pr}(\text{Reject }H_t|H_t)\\
    &=\pr_{\boldsymbol{\alpha}_1,\dots,\boldsymbol{\alpha}_d,\boldsymbol{X}}\left(\argmax_{1\leq k\leq d}\mathcal{L}(\boldsymbol{X};\boldsymbol{\alpha}_k,\boldsymbol{\theta})\neq t|H_t\right)\\
    &\leq \sum_{k\neq t}\pr_{\boldsymbol{\alpha}_1,\dots,\boldsymbol{\alpha}_d,\boldsymbol{X}}(\mathcal{L}(\boldsymbol{X};\boldsymbol{\alpha}_k,\boldsymbol{\theta})>\mathcal{L}(\boldsymbol{X};\boldsymbol{\alpha}_t,\boldsymbol{\theta})|H_t)\\
    &=\sum_{k\neq t}\E_{\boldsymbol{\alpha}_1,\dots,\boldsymbol{\alpha}_d}\Phi\left(-\frac{\sqrt{\kappa N_S}}{N_B}\|\Re((\boldsymbol{\alpha}_t^*-\boldsymbol{\alpha}_k^*)\circ e^{-i\boldsymbol{\theta}})\|\right),
\end{align}
where the third line follows from the union bound, which is asymptotically tight in the low-error regime.
Here, $\|\cdot\|$ denotes the Euclidean norm and $\Phi(\cdot)$ the cumulative distribution function of the standard normal distribution, defined as
\begin{align}
    \Phi(x)\coleq\int_{-\infty}^x\frac{1}{\sqrt{2\pi}}e^{-s^2/2}ds.
\end{align}
Using the asymptotically tight upper bound $\Phi(-x)\leq e^{-x^2/2}/2$, we obtain
\begin{align}
    &\E_{\boldsymbol{\alpha}_1,\dots,\boldsymbol{\alpha}_d}\Phi\left(-\frac{\sqrt{\kappa N_S}}{N_B}\|\Re((\boldsymbol{\alpha}_t^*-\boldsymbol{\alpha}_k^*)\circ e^{-i\boldsymbol{\theta}})\|\right)\\
    &\quad\leq\frac{1}{2}\E_{\boldsymbol{\alpha}_1,\dots,\boldsymbol{\alpha}_d}\exp\left(-\frac{\kappa N_S}{2N_B^2}\|\Re((\boldsymbol{\alpha}_t^*-\boldsymbol{\alpha}_k^*)\circ e^{-i\boldsymbol{\theta}})\|^2\right)\\
    &\quad=\frac{1}{2}\E_{\boldsymbol{\alpha}_1,\dots,\boldsymbol{\alpha}_d}\exp\left(-\frac{\kappa N_S}{2N_B^2}\sum_{l=1}^M(\Re((\alpha_{t,l}^*-\alpha_{k,l}^*) e^{-i\theta_l}))^2\right)\\
    &\quad=\frac{1}{2}\left(\E_{{\alpha}_{1,l},\dots,{\alpha}_{d,l}}\exp\left(-\frac{\kappa N_S}{2N_B^2}(\Re((\alpha_{t,l}^*-\alpha_{k,l}^*) e^{-i\theta_l}))^2\right)\right)^M
\end{align}
for an arbitrary $l\in[M]$.
To simplify the expression, we derive the approximation of the exponential term on the RHS.
Defining the random variable $Y\coleq \frac{\kappa N_S}{2N_B^2}(\Re((\alpha_{t,l}^*-\alpha_{k,l}^*) e^{-i\theta_l}))^2$ and omitting subscripts in the expectation for simplicity, we have
\begin{align}
    \mathrm{RHS}&=\frac{1}{2}(\E\exp(-Y))^M\\
    &=\frac{1}{2}\exp(M\log(\E\exp(-Y)))\\
    &=\frac{1}{2}\exp(M\log(\E(1-Y+Y^2/2-\cdots)))\\
    &=\frac{1}{2}\exp(M\log(1-\E Y+\mathcal{O}((\E Y)^2)))\\
    &\approx\frac{1}{2}\exp(M(\log\exp(-\E Y)))\\
    &=\frac{1}{2}\exp(-M\E Y),
\end{align}
where the fourth line follows from $\mathcal{O}((\E Y)^2))=\mathcal{O}((\kappa N_S/N_B)^2)\ll 1$.
Collecting these, we obtain the asymptotically tight upper bound of the error probability as
\begin{align}
    P_{\text{error}}
    &\leq\sum_{k\neq t}\frac{1}{2}\exp\left(-\frac{\kappa N_S M}{2N_B^2}\E_{{\alpha}_{1,l},\dots,{\alpha}_{d,l}}(\Re((\alpha_{t,l}^*-\alpha_{k,l}^*) e^{-i\theta_l}))^2\right)\\
    &=\frac{d-1}{2}\exp\left(-\frac{\kappa N_S M}{2N_B^2}\E_{{\alpha}_{1,l},\dots,{\alpha}_{d,l}}(\Re((\alpha_{t,l}^*-\alpha_{k,l}^*) e^{-i\theta_l}))^2\right)\\
    &\sim\exp(-\xi M),
\end{align}
with the error exponent
\begin{align}
    \xi=\frac{\kappa N_S}{2N_B^2}\E_{{\alpha}_{1,l},\dots,{\alpha}_{d,l}}(\Re((\alpha_{t,l}^*-\alpha_{k,l}^*) e^{-i\theta_l}))^2.
\end{align}
Here, the second line follows from the fact that $\theta_l$ is symmetric with respect to $\alpha_{1,l},\dots,\alpha_{d,l}$, which also allows us to choose $k\in[d]\backslash\{t\}$ arbitrarily in defining the error exponent.
For notational simplicity, we omit the subscript $l$ and write
\begin{align}
    \xi=\frac{\kappa N_S}{2N_B^2}\E_{\alpha_1,\dots,\alpha_d}(\Re((\alpha_{t}^*-\alpha_{k}^*) e^{-i\theta}))^2.
    \label{eq:xi_c}
\end{align}
for $\alpha_1,\dots,\alpha_d\sim \mathcal{CN}(0,N_B)$, with a slight abuse of notation.

We now complete the proof by deriving the error exponent $\xi$ in a closed form.
We begin by rewriting the error exponent in terms of real Gaussian variables.
Following the isomorphism $\mathbb{C}\cong \mathbb{R}^2$, we adopt the notations
\begin{align}
    e^{i\theta}&\cong \boldsymbol{u}\coleq (\cos\theta, \sin\theta)^T,\\
    \alpha_k^*&\cong \boldsymbol{v}_k \coleq (\Re(\alpha_k^*), \Im(\alpha_k^*))^T,
\end{align}
for $k\in [d]$, where $\boldsymbol{v}_k\sim \mathcal{N}(\boldsymbol{0}_2, (N_B/2) I_2)$.
Then, the error exponent can be written as
\begin{align}
    \xi
    &=\frac{\kappa N_S}{2N_B^2}\E_{\boldsymbol{v}_1,\dots,\boldsymbol{v}_d}((\boldsymbol{v}_t-\boldsymbol{v}_k)^T\boldsymbol{u})^2\\
    &=\frac{\kappa N_S}{2N_B^2}\E_{\boldsymbol{v}_1,\dots,\boldsymbol{v}_d}\boldsymbol{u}^T(\boldsymbol{v}_t-\boldsymbol{v}_k)(\boldsymbol{v}_t-\boldsymbol{v}_k)^T\boldsymbol{u}
\end{align}
with $\boldsymbol{v}_1,\dots,\boldsymbol{v}_d\sim \mathcal{N}(\boldsymbol{0}_2,(N_B/2)I_2)$ for an arbitrary $k\in [d]\backslash\{t\}$, where $\boldsymbol{0}_2$ denotes the 2-dimensional zero vector.
As noted earlier, the symmetry of $\theta$ ensures that $\xi$ is independent of the choice of $t$ and $k$.
Thus, by averaging over all $t$ and $k$, we obtain
\begin{align}
    \xi
    &=\frac{\kappa N_S}{2N_B^2}\E_{\boldsymbol{v}_1,\dots,\boldsymbol{v}_d}\frac{1}{d(d-1)} \sum_{t,k=1}^d \boldsymbol{u}^T(\boldsymbol{v}_t-\boldsymbol{v}_k)(\boldsymbol{v}_t-\boldsymbol{v}_k)^T\boldsymbol{u} \label{eq:exp_v}\\
    &=\frac{\kappa N_S}{2N_B^2}\E_{\boldsymbol{v}_1,\dots,\boldsymbol{v}_d}\frac{2}{d-1}\boldsymbol{u}^T\left(\sum_{t=1}^d (\boldsymbol{v}_t-\boldsymbol{\bar{v}})(\boldsymbol{v}_t-\boldsymbol{\bar{v}})^T\right)\boldsymbol{u}\\
    &=\frac{\kappa N_S}{(d-1)N_B^2}\E_{\boldsymbol{v}_1,\dots,\boldsymbol{v}_d}\boldsymbol{u}^TS\boldsymbol{u}
\end{align}
for 
\begin{align}
    S\coleq \sum_{t=1}^d (\boldsymbol{v}_t-\boldsymbol{\bar{v}})(\boldsymbol{v}_t-\boldsymbol{\bar{v}})^T,
\end{align}
where $\boldsymbol{\bar{v}}\coleq \sum_{t=1}^d\boldsymbol{v}_t/d$ denotes the mean of the $\boldsymbol{v}_k$ and the second line follows from a simple algebraic manipulation~\cite{muirhead2010aspects}.
From the given construction of $\theta$, $\boldsymbol{u}$ is the first principal component of $S$.
Consequently, $\boldsymbol{u}$ is the eigenvector of $S$ corresponding to its maximum eigenvalue, which leads to the error exponent
\begin{align}
    \xi=\frac{\kappa N_S}{(d-1)N_B^2}\E\lambda_{\max}(S).
\end{align}
Here, $\lambda_{\max}(\cdot)$ denotes the maximum eigenvalue of a matrix, and the subscript of the expectation operator is omitted for notational simplicity.

Now, our goal is reduced to finding the expectation of the maximum eigenvalue of a random matrix $S$ for $\boldsymbol{v}_1,\dots,\boldsymbol{v}_d\sim \mathcal{N}(\boldsymbol{0}_2,(N_B/2)I_2)$.
Such a random matrix is often referred to as a scatter matrix, since it serves as an estimator of the covariance matrix of scattered Gaussian random variables $\boldsymbol{v}_1,\dots,\boldsymbol{v}_d$.
A standard result is that the normalized scatter matrix $\bar{S}\coleq S/(N_B/2)$ follows the Wishart distribution, $\bar{S}\sim W(2, d-1)$~\cite{muirhead2010aspects}.
Here, the Wishart distribution $W(m, n)$ is the distribution of an $m\times m$ random matrix $GG^T$, where $G$ is an $m\times n$ random matrix with \textit{i.i.d.} entries $\mathcal{N}(0, 1)$.
Based on this property, we obtain the expectation of the maximum eigenvalue $\lambda_{\max}(\bar{S})$ by using the spectral properties of the Wishart distribution.
Denoting $n=d-1$ and $\lambda_1\geq \lambda_2\geq0$ as the ordered eigenvalues of $\bar{S}\sim W(2, n)$, the joint density of the eigenvalues is given as
\begin{align}
f(\lambda_1, \lambda_2) = A_n\, e^{-\frac{\lambda_1+\lambda_2}{2}}\, (\lambda_1 \lambda_2)^{\frac{n - 3}{2}} (\lambda_1 - \lambda_2)
\end{align}
with the normalization constant
\begin{align}
A_n = \frac{2^{-n}\pi^{1/2}}{\Gamma(n/2)\, \Gamma((n-1)/2)},
\end{align}
where $\Gamma(z)\coleq\int_0^\infty t^{z-1}e^{-t}\diff t$ is the Gamma function~\cite{james1964distributions, skinner1985introduction}.
We then decompose the expectation as
\begin{align}
\E\lambda_{\max}(\bar{S})\equiv \E \lambda_1 = \frac{1}{2}\E (\lambda_1 + \lambda_2)+ \frac{1}{2} \E(\lambda_1 - \lambda_2).
\end{align}
For the first term, we have $\E(\lambda_1 + \lambda_2) = \E\mathrm{Tr}(\bar{S}) = \mathrm{Tr}(\E\bar{S}) = 2n=2(d-1)$, since the diagonal elements of $\bar{S}$ follow a $\chi^2_n$-distribution.
Thus,
\begin{align}
\E\lambda_{\max}(\bar{S}) 
&= d-1+\frac{1}{2}\E(\lambda_1 - \lambda_2),
\end{align}
leaving only the second term to be evaluated.
Carrying out the integral, we obtain
\begin{align}
\E(\lambda_1 - \lambda_2) = A_n \int_0^\infty \int_0^{\lambda_1}  e^{-\frac{\lambda_1+\lambda_2}{2}} (\lambda_1 \lambda_2)^{\frac{n-3}{2}}  (\lambda_1 - \lambda_2)^2\diff\lambda_2 \diff\lambda_1.
\end{align}
After changing variables $\lambda_2 = r \lambda_1$ with $r \in [0,1]$, the integral becomes
\begin{align}
&A_n\int_0^\infty\int_0^1 e^{-\frac{\lambda_1(1+r)}{2}} \lambda_1^n r^{\frac{n-3}{2}} (1 - r)^2  \diff r\diff\lambda_1\\
&\quad=A_n\int_0^1r^{\frac{n-3}{2}}(1-r)^2\int_0^\infty e^{-\frac{\lambda_1(1+r)}{2}} \lambda_1^n \diff\lambda_1 \diff r\\
&\quad=A_n\int_0^1r^{\frac{n-3}{2}}(1-r)^2  \left(\frac{2}{1+r}\right)^{n+1}\Gamma(n+1) \diff r\\
&\quad=2^{n+1}\Gamma(n+1)A_n\int_0^1 \frac{r^{\frac{n-3}{2}}(1-r)^2}{(1+r)^{n+1}} \diff r.
\end{align}
Changing variables further with $s=(1-r)/(1+r)$, we have $r=(1-s)/(1+s)$ and $\diff r=-2\diff s/(1+s)^2$, which transforms the integral into
\begin{align}
    &2^{n+1}\Gamma(n+1)A_n\int_1^0\left(\frac{1-s}{1+s}\right)^{\frac{n-3}{2}}\left(\frac{2s}{1+s}\right)^2\left(\frac{1+s}{2}\right)^{n+1}\frac{-2}{(1+s)^2}\diff s\\
    &\quad = 8\Gamma(n+1) A_n\int_0^1(1-s^2)^{\frac{n-3}{2}}s^2 \diff s
\end{align}
Finally, substituting $t=s^2$ with $s=\sqrt{t}$ and $\diff s=\diff t/(2\sqrt{t})$ yields
\begin{align}
    &4\Gamma(n+1) A_n\int_0^1(1-t)^{\frac{n-3}{2}}t^{\frac{1}{2}} \diff t\\
    &\quad = 4\Gamma(n+1) A_n \mathrm{B}\left(\frac{3}{2},\frac{n-1}{2}\right)\\
    &\quad = 4\Gamma(n+1) \frac{2^{-n}\pi^{1/2}}{\Gamma(n/2)\, \Gamma((n-1)/2)} \frac{\Gamma(3/2)\Gamma((n-1)/2)}{\Gamma((n+2)/2)}\\
    &\quad=2^{-(n-1)}\pi\frac{\Gamma(n+1)}{\Gamma(n/2)\Gamma(n/2+1)}\\
    &\quad=2^{-(d-2)}\pi\frac{\Gamma(d)}{\Gamma((d-1)/2)\Gamma((d+1)/2)}.
\end{align}
Combining everything, we obtain
\begin{align}
    \xi&=
    \frac{\kappa N_S}{(d-1)N_B^2}\E\lambda_{\max}(S)\\
    &=
    \frac{\kappa N_S}{2(d-1)N_B}\E\lambda_{\max}(\bar{S})\\
    &=\frac{\kappa N_S}{2(d-1)N_B}\left(d-1+\frac{1}{2}\E(\lambda_1-
    \lambda_2)\right)\\
    &=\frac{\kappa N_S}{2(d-1)N_B}\left(d-1+2^{-(d-1)}\pi\frac{\Gamma(d)}{\Gamma((d-1)/2)\Gamma((d+1)/2)}\right)\\
    &=\frac{\kappa N_S}{2N_B}\left(1+\frac{2^{-(d-1)}\pi\Gamma(d-1)}{\Gamma((d-1)/2)\Gamma((d+1)/2)}\right)\\
    &=\frac{\kappa N_S}{2N_B}\left(1+\frac{\sqrt{\pi}\Gamma(d/2)}{2\Gamma((d+1)/2)}\right)\\
    &=\frac{\kappa N_S}{2N_B}\left(1+\frac{\Gamma(1/2)\Gamma(d/2)}{2\Gamma((d+1)/2)}\right)\\
    &=\frac{\kappa N_S}{2N_B}\left(1+\frac{\mathrm{B}(d/2,1/2)}{2}\right),\label{eq:exp}
\end{align}
where $\mathrm{B}(z_1,z_2)\coleq\Gamma(z_1)\Gamma(z_2)/\Gamma(z_1+z_2)$ is the Beta function.
The sixth line follows from the Legendre duplication formula $\Gamma(z)\Gamma(z+1/2)=2^{1-2z}\sqrt{\pi}\Gamma(2z)$.
Moreover, one can approximate the error exponent in the regime $d\gg 1$ as
\begin{align}
    \xi\approx\frac{\kappa N_S}{2N_B}\left(1+\sqrt{\frac{\pi}{2d}}\right)
\end{align}
using the known asymptotic $\mathrm{B}(z_1,z_2)\approx\Gamma(z_2)z_1^{-z_2}$ for a large $z_1$ with fixed $z_2$.

Additionally, we show that the error exponent achieves a 3 dB advantage with $\xi = 2\xi_{\mathrm{CTR}}$ for $d = 2$ in a simpler way. 
In this case, Eq.~\eqref{eq:exp_v} becomes
\begin{align}
    \xi&=\frac{\kappa N_S}{2N_B^2}\E_{\boldsymbol{v}_1,\boldsymbol{v}_2}\boldsymbol{u}^T(\boldsymbol{v}_1-\boldsymbol{v}_2)(\boldsymbol{v}_1-\boldsymbol{v}_2)^T\boldsymbol{u}\\
    &=\frac{\kappa N_S}{2N_B^2}\E_{\boldsymbol{v}_1,\boldsymbol{v}_2}\lambda_{\max}((\boldsymbol{v}_1-\boldsymbol{v}_2)(\boldsymbol{v}_1-\boldsymbol{v}_2)^T)
\end{align}
for $\boldsymbol{v}_1, \boldsymbol{v}_2\sim\mathcal{N}(\boldsymbol{0}_2,(N_B/2)I_2)$.
Since $(\boldsymbol{v}_1-\boldsymbol{v}_2)(\boldsymbol{v}_1-\boldsymbol{v}_2)^T$ is a rank-1 matrix, we can write the error exponent as
\begin{align}
    \xi
    &=\frac{\kappa N_S}{2N_B^2}\E_{\boldsymbol{v}_1,\boldsymbol{v}_2}\tr((\boldsymbol{v}_1-\boldsymbol{v}_2)(\boldsymbol{v}_1-\boldsymbol{v}_2)^T)\\
    &=\frac{\kappa N_S}{2N_B^2}\E_{\boldsymbol{v}_1,\boldsymbol{v}_2}(\boldsymbol{v}_1-\boldsymbol{v}_2)^T(\boldsymbol{v}_1-\boldsymbol{v}_2)\\
    &=\frac{\kappa N_S}{2N_B}\E_{\boldsymbol{v}_1,\boldsymbol{v}_2}\left(\frac{\boldsymbol{v}_1-\boldsymbol{v}_2}{\sqrt{N_B}}\right)^T\left(\frac{\boldsymbol{v}_1-\boldsymbol{v}_2}{\sqrt{N_B}}\right)\\
    &=\frac{\kappa N_S}{N_B}\\
    &=2\xi_{\mathrm{CTR}},
\end{align}
where the fourth line follows from the fact that $((\boldsymbol{v}_1-\boldsymbol{v}_2)/\sqrt{N_B})^T((\boldsymbol{v}_1-\boldsymbol{v}_2)/\sqrt{N_B})\sim\chi_2^2$.

\endgroup

\end{document}